# ELECTROWEAK PHYSICS RESULTS FROM THE DØ EXPERIMENT AT FERMILAB[1]

JOHN ELLISON

*Physics Department, University of California,*

*Riverside, CA 92521, USA*

for the DØ Collaboration

ABSTRACT

We present electroweak physics results from the DØ experiment using data from $p\bar{p}$ collisions at $\sqrt{s} = 1.8$ TeV, concentrating on *single W and Z production cross sections* and the production of electroweak gauge boson *pairs*. The inclusive cross sections times branching ratios for *W* and *Z* production have been measured and are used to determine the inclusive width of the *W* boson. Direct tests of the $WW\gamma$ and $WWZ$ trilinear gauge boson couplings are derived from the study of diboson production. These measurements test the non-abelian self couplings of the *W*, *Z* and photon, one of the most direct consequences of the $SU(2)_L \times U(1)_Y$ gauge symmetry. We present measurements of the $WW\gamma$ coupling based on $p\bar{p} \to l\nu\gamma + X$ ($l = e, \mu$) and limits on anomalous $WWZ$ and $WW\gamma$ couplings based on searches for $p\bar{p} \to W^+W^- + X \to ll'\nu\bar{\nu} + X$ and $p\bar{p} \to WW/WZ + X \to e\nu + \text{jets} + X$. We also present limits on the $ZZ\gamma$ and $Z\gamma\gamma$ couplings based on $p\bar{p} \to l\bar{l}\gamma + X$ events.

## 1. Introduction

The DØ detector at the Fermilab Tevatron $p\bar{p}$ collider collected data corresponding to an integrated luminosity of 15 pb$^{-1}$ in its first data taking run ("Run 1a") from 1992–93 and 88 pb$^{-1}$ in its second run ("Run 1b") during 1994–95. In this paper we present some results on electroweak physics from the analysis of a subset of this data (mainly that from Run 1a).

The experimental study of the physics of electroweak gauge bosons began in 1983 with the discovery of the *W* and *Z* bosons at the CERN $Sp\bar{p}S$ [1]. The production of *W* and *Z* bosons in much larger numbers has subsequently been possible at LEP, SLC and the Tevatron. The Tevatron data allows refined measurements of the *W* boson properties, such as its mass and width, with improved precision, and *new* studies of the physics of electroweak boson pair production. In this paper we concentrate on the latter topic (sections 3–6). As a prelude we present the measurements of the $W$ and $Z$ production cross sections (section 2).

---

[1] Presented at the X$^{th}$ International Workshop on High Energy Physics and Quantum Field Theory, Zvenigorod, Russia, 20-26 September 1995.

## 2. *W* and *Z* Production and the Inclusive Width of the *W* Boson

The measurement of the production cross-sections times branching ratios ($\sigma \cdot B$) for the $W$ and $Z$ bosons allows a determination of the width of the $W$ boson and a comparison of $W$ and $Z$ boson production with QCD predictions.

The first DØ measurement of $\sigma \cdot B$ was performed using data collected in Run 1a corresponding to a total integrated luminosity of approximately 12 pb$^{-1}$ for the triggers used for $W$ and $Z$ event filtering.

Measurement of the cross sections were made using the $e, \mu$ decay mode channels. Measurements using $\tau$ decays and $qq$ decays are experimentally challenging due to the low efficiency of $\tau$ detection and very high QCD background in the dijet channel. The decay channels used in this analysis are $W \rightarrow e\nu$, $W \rightarrow \mu\nu$, $Z \rightarrow e^+e^-$ and $Z \rightarrow \mu^+\mu^-$. Events were selected from single lepton triggers and offline were required to contain a high $p_T$ isolated lepton plus missing transverse energy ($W$ events) or two high $p_T$ isolated leptons ($Z$ events), one passing the standard identification cuts and the second passing less stringent "loose" cuts. The selection criteria are summarized in Table 1. Details of the DØ detector and particle identification criteria are given in [2,3].

|  | $W \rightarrow e\nu$ | $Z \rightarrow e^+e^-$ | $W \rightarrow \mu\nu$ | $Z \rightarrow \mu^+\mu^-$ |
|---|---|---|---|---|
| Selection Cuts: | | | | |
| $E_T^{l1}$ (GeV) | 25 | 25 | 20 | 20 |
| $E_T^{l2}$ or $\not{E}_T$ (GeV) | 25 | 25 | 20 | 15 |
| Lepton $\eta$ | $|\eta|<1.1$ or $1.5<|\eta|<2.5$ | | $|\eta|<1.0$ | |
| No. Candidates | 10338 | 775 | 1665 | 77 |
| Backgrounds (%): | | | | |
| $Z \rightarrow ll$ | 0.6 ± 0.1 | – | 7.3 ± 0.5 | 0.7 ±0.2 |
| $W \rightarrow \tau\nu$ | 1.8 ± 0.1 | – | 5.9 ± 0.5 | – |
| QCD | 3.3 ± 0.5 | 2.8 ± 1.4 | 5.1 ± 0.8 | 2.6 ± 0.8 |
| Cosmic | – | – | 3.8 ± 1.6 | 5.1 ± 3.6 |
| Drell-Yan | – | 1.2 ± 0.1 | – | 1.7 ± 0.3 |
| Total | 5.7 ± 0.5 | 4.0 ± 1.4 | 22.1 ±1.9 | 10.1 ± 3.7 |
| Acceptance (%) | 46.0 ± 0.6 | 36.3 ± 0.4 | 24.8 ± 0.7 | 6.5 ±0.4 |
| Efficiency (%) | 70.4 ± 1.7 | 73.6 ± 2.4 | 21.9 ± 2.6 | 52.7 ± 4.9 |
| $\int Ldt$ (pb$^{-1}$) | 12.8 ± 0.7 | 12.8 ± 0.7 | 11.4 ± 0.6 | 11.4 ± 0.6 |
| $\sigma \cdot B$ (nb) | 2.36 | 0.218 | 2.09 | 0.178 |
| stat error | ±0.02 | ±0.008 | ±0.06 | ±0.022 |
| syst error | ±0.07 | ±0.008 | ±0.22 | ±0.021 |
| lum error | ±0.13 | ±0.012 | ±0.11 | ±0.009 |

Table 1. Summary of analysis and production cross section times branching ratio results for *W* and *Z* bosons.

Figure 1 shows the transverse mass spectra for $W \to e\nu$ and $W \to \mu\nu$ events and the invariant mass spectra for $Z \to e^+e^-$ and $Z \to \mu^+\mu^-$ events (data points) together with the background estimates (shaded histograms) and the sum of the background distributions and a Monte Carlo simulation of the signal, using GEANT [4] to model the DØ detector. The backgrounds, ranging between 4% to 22% of the signal, depending on the channel, are listed in Table 1.

The values of $\sigma \cdot B$ were calculated by subtracting the background from the number of observed events and dividing by the acceptance, efficiency and integrated luminosity. The results are shown in Table 1 and are plotted in Fig. 2, together with the CDF results from the same collider run and *preliminary* DØ results using $\approx 30$ pb$^{-1}$ of data from the most recent data taking run (Run 1b). The measurements agree well with the order $\alpha_s^2$ QCD predictions of $\sigma_W \cdot B(W \to l\nu) = 2.42^{+0.13}_{-0.11}$ nb and $\sigma_Z \cdot B(Z \to ll) = 0.226^{+0.011}_{-0.009}$ nb, shown as the shaded bands in Fig. 2. The CTEQ2M parton distribution functions [5] were used to calculate the central values.

The inclusive width of the W boson is calculated using the measured ratio ($R$) of the $W$ and $Z$ $\sigma \cdot B$ values:

$$R = \frac{\sigma_W \cdot B(W \to l\nu)}{\sigma_Z \cdot B(Z \to ll)} \quad \text{with} \quad B(W \to l\nu) = \frac{\Gamma(W \to l\nu)}{\Gamma(W)}.$$

Combining the DØ Run 1a electron and muon channel data yields $R_{e+\mu} = 10.89 \pm 0.49 (\text{stat} \oplus \text{syst})$. Many common sources of error cancel in $R$, including the uncertainty in the integrated luminosity and parts of the errors in the acceptance and event selection efficiency. The theoretical calculation $\sigma_W / \sigma_Z = 3.33 \pm 0.03$ [3,6] and the LEP measurement $B(Z \to ll) = (3.367 \pm 0.006)\%$ [7] are then used to obtain $B(W \to l\nu) = (11.02 \pm 0.50)\%$. We combine this measurement of $B(W \to l\nu)$ with a theoretical calculation [3,8] of $\Gamma(W \to l\nu) = 225.2 \pm 1.5$ MeV to obtain an indirect measurement of the $W$ inclusive width:

$$\Gamma(W) = 2.044 \pm 0.093 \text{ GeV}.$$

Figure 3 summarizes the measurements of $\Gamma(W)$ to date. The world average of $2.062 \pm 0.059$ GeV can be compared with the Standard Model (SM) prediction of $2.077 \pm 0.014$ GeV [3,8] to set limits on non-standard decays of the $W$. At the 95% confidence level, the upper limit on the width due to non-standard decays (*e.g.* decays to heavy quarks or to SUSY particles such as charginos and neutralinos) is 109 MeV.

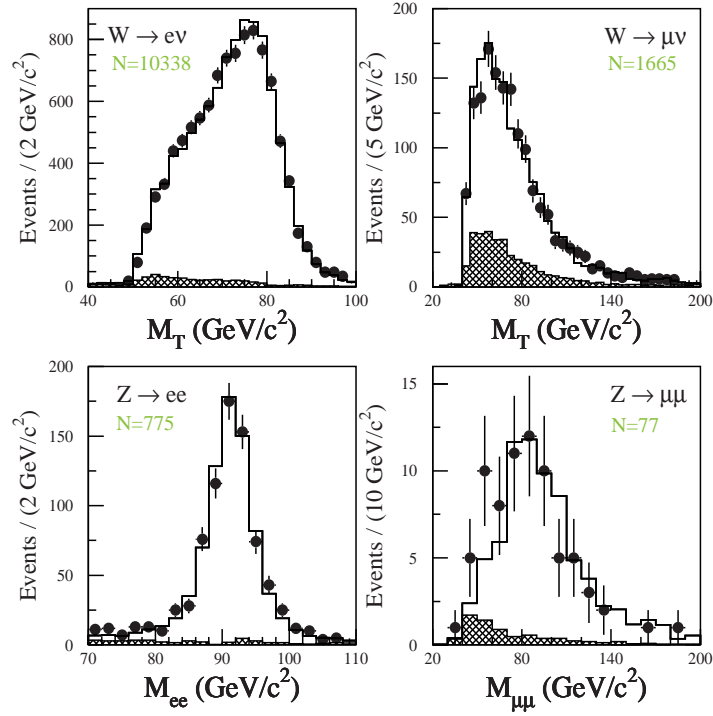

Fig. 1. Transverse mass and invariant mass distributions for the *W* and *Z* candidate events (points). The shaded histograms show the total estimated backgrounds and the open histogram is the sum of the SM signal plus the total background.

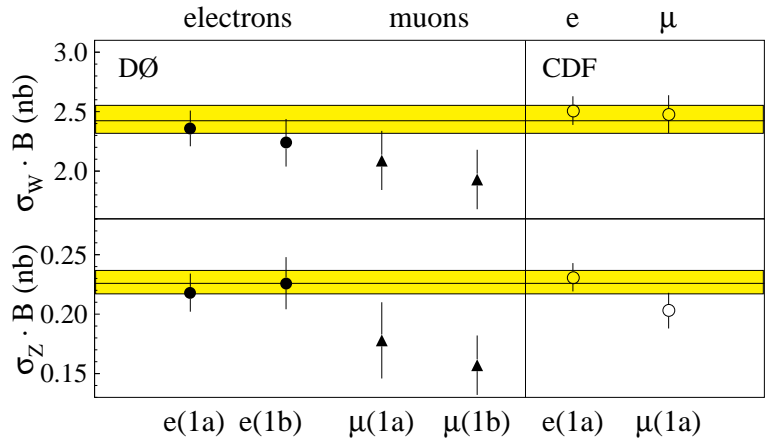

Fig. 2. Measurements of the *W* and *Z* inclusive cross-sections times branching ratios from DØ (solid points). Also shown are the measurements from CDF (open points) and the SM theoretical predictions (shaded bands show central values and $\pm 1\sigma$ uncertainties). The central values were obtained using the CTEQ2M parton distribution functions [5].

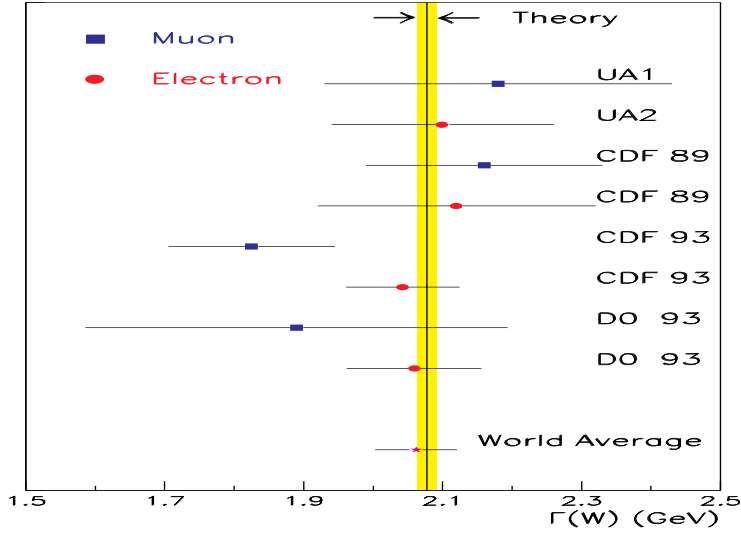

Fig. 3.   Measurements of the *W* width from the Tevatron and CERN compared with the SM theoretical prediction (band shows $\pm 1\sigma$ uncertainties).

## 3. Diboson Production and the Gauge Boson Couplings

The production of electroweak boson pairs at hadron colliders is particularly interesting since these processes probe the nature of the non-Abelian gauge boson self-couplings of the Standard Model (*i.e.* the *WW*$\gamma$, *WWZ* couplings). These processes are also sensitive to deviations from the tree level SM couplings, which may arise due to compositeness of the *W* and *Z* bosons or loop corrections to the vertex factors as shown in the examples of Fig. 4. These effects may also result in non-zero *ZZ*$\gamma$ and *Z*$\gamma\gamma$ couplings which can be tested via *Z*$\gamma$ production. In the SM these couplings are zero since the *Z* is a neutral particle and does not couple to the neutral photon[2].

To test the agreement with the SM and to evaluate the sensitivity to anomalous couplings we use a phenomenological effective Lagrangian for the *WWV* (*V*=$\gamma$,*Z*) vertex, assuming electromagnetic gauge invariance, and invariance under Lorentz and CP transformations[3] [9]:

$$i\mathcal{L}_{eff}^{WWV} = g_{WWV}\left[ g_1^V(W_{\mu\nu}^\dagger W^\mu - W^{\dagger\mu}W_{\mu\nu})V^\nu + \kappa_V W_\mu^\dagger W_\nu V^{\mu\nu} + \frac{\lambda_V}{m_W^2}W_{\rho\mu}^\dagger W_\nu^\mu V^{\nu\rho}\right]$$

where $W^\mu$ and $V^\mu$ are the *W* and *V* fields respectively, $g_{WW\gamma} = -e$, $g_{WWZ} = -e\cot\theta_W$ and $g_1^V, \kappa_V$, and $\lambda_V$ are dimensionless coupling parameters. In the SM at tree level $g_1^V = 1, \kappa_V = 1$, and $\lambda_V = 0$. Assuming that the coupling parameters for the *WW*$\gamma$ and *WWZ* vertices are equal, only two parameters remain:

---

[2] Both the photon and *Z* have zero electric charge and zero weak isospin.
[3] CP violating operators may also be introduced in the Lagrangian, but for simplicity, we do not consider them here.

$$\kappa_\gamma = \kappa_Z = \kappa \quad (\text{or } \Delta\kappa \equiv \kappa - 1)$$
$$\lambda_\gamma = \lambda_Z = \lambda.$$

The coupling parameters are related to the lowest order terms in a multipole expansion of photon interactions with the $W$ boson, e.g. the $W$ magnetic dipole moment is given by $\mu_W = (e/2m_W)(g_1^\gamma + \kappa_\gamma + \lambda_\gamma)$.

A similar formalism is used to describe the $Z\gamma V$ ($V=\gamma,Z$) vertices. It is conventional to parametrize the couplings in terms of the $Z\gamma V$ vertex functions [9]:

$$\Gamma_{Z\gamma V}^{\alpha\beta\mu}(q_1, q_2, P) = \frac{P^2 - m_V^2}{m_Z^2}\left[ h_3^V \varepsilon^{\mu\alpha\beta\rho} q_{2\rho} + \frac{h_4^V}{m_Z^2} P^\alpha \varepsilon^{\mu\beta\rho\sigma} P_\rho q_{2\sigma} \right]$$

where $h_3^V, h_4^V$ are dimensionless coupling parameters, $q_1$, $q_2$ and $P$ are the 4-momenta of the $Z$, $\gamma$ and $V$ respectively. In the SM at tree level (no coupling of the $\gamma$ to the $Z$) $h_3^V = 0, h_4^V = 0$.

All interaction Lagrangians with constant anomalous couplings violate unitarity at high energy and so the coupling parameters must be modified to include form factors, e.g.

$$\Delta\kappa(q^2) = \frac{\Delta\kappa^0}{(1 + q^2/\Lambda^2)^n}$$

where $\Delta\kappa^0$ = value of coupling parameter at $q^2 = 0^4$, $q^2 = \hat{s}$ = square of the invariant mass of the $W\gamma$ system, $n = 2$ for a dipole form factor and $\Lambda$ = form factor scale (a function of the scale of new physics). We take $n=2$ for $\Delta\kappa$ and $\lambda$. For $h_3^V$ we take $n=3$ and for $h_4^V$ $n=4$, which ensures that unitarity is satisfied and that both terms have the same high energy behavior [10]. Note that in the SM delicate cancellations (such as those between the first three Feynman graphs in Fig. 5) ensure that unitarity is satisfied. Constant anomalous couplings would destroy these cancellations leading to violation of unitarity.

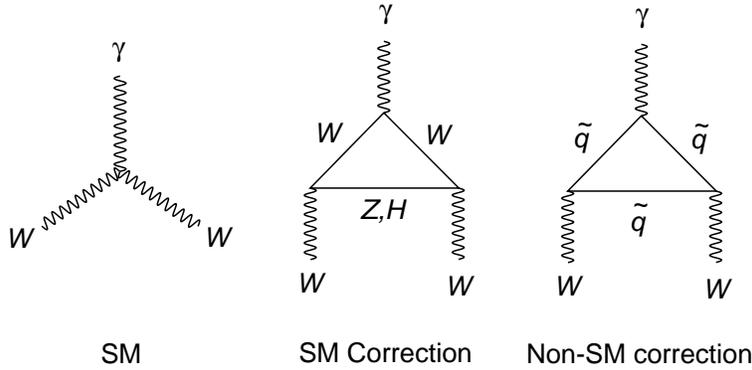

Fig. 4. Standard Model $WW\gamma$ vertex (left) and examples of loop corrections leading to deviations from the tree-level SM vertex functions.

---

[4] The notations for the other couplings defined at $q^2 = 0$ are $\lambda^0$, $h_{30}^V$ and $h_{40}^V$.

## 4. $W\gamma$ Production and the $WW\gamma$ Coupling

$W\gamma$ production at the Tevatron can be used to probe the $WW\gamma$ coupling [11]. The Feynman diagrams for the processes contributing to $p\bar{p} \to l\nu\gamma + X$ are shown in Fig. 5. The u- and t-channel diagrams correspond to photon bremsstrahlung from an initial state quark, whereas the s-channel diagram is sensitive to the $WW\gamma$ vertex. Events in which a photon is radiated from the final state lepton from single $W$ decay also result in the same $l\nu\gamma$ final state. These radiative decay events can be suppressed by imposing a cut on the photon-lepton angular separation and further by requiring the lepton-photon-neutrino transverse cluster mass[5] $M_T(l\gamma;\nu)$ to be $> M_W$. Non-SM values of the $WW\gamma$ coupling (Fig. 5 (c)) result in an increase in the total cross-section and an enhacement of events with high-$E_T$ photons. This is the experimental signature which we use to test for anomalous couplings.

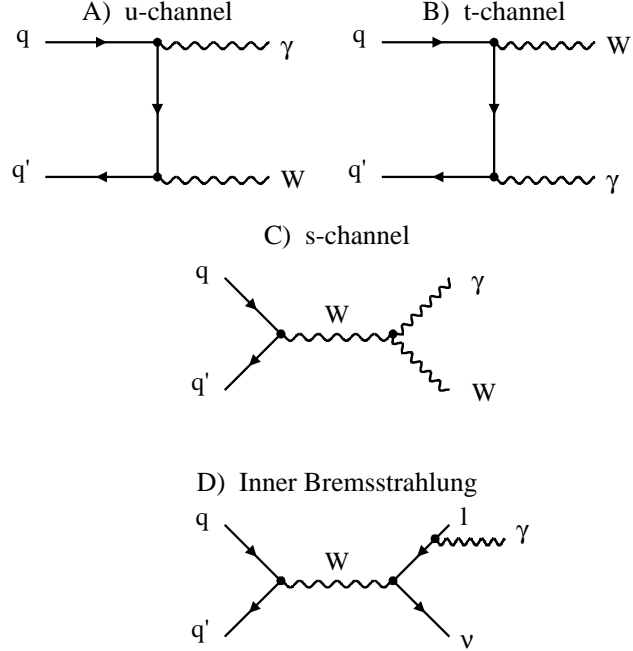

Fig. 5.  Feynman diagrams contributing to the process $p\bar{p} \to l\nu\gamma$.

We selected a sample of $W(e\nu)\gamma$ candidates from events passing a trigger which requires an isolated EM cluster with $E_T > 20$ GeV and missing transverse energy ($\not{E}_T$) > 20 GeV. This EM cluster was required to be within the fiducial region of the calorimeter, $|\eta| < 1.1$, and at least 0.01 radians away from the azimuthal boundaries between the 32 EM modules in the central calorimeter, or within $1.5 < |\eta| < 2.5$ in the end calorimeters. The offline kinematic selection was made requiring $E_T^e > 25$ GeV, $\not{E}_T > 25$ GeV and a transverse mass $M_T > 40$ GeV. The electron was required to satisfy electron identification criteria as described in [12].

---

[5] The transverse cluster mass is defined by $M_T^2(l\gamma;\nu) = \left[\left(m_{l\gamma}^2 + |\mathbf{E}_T^\gamma + \mathbf{E}_T^l|^2\right)^{1/2} + \slashed{E}_T\right]^2 - \left|\mathbf{E}_T^\gamma + \mathbf{E}_T^l + \slashed{\mathbf{E}}_T\right|^2$ where $m_{l\gamma}$ is the photon–lepton invariant mass, $\mathbf{E}_T^\gamma$ and $\mathbf{E}_T^l$ are the photon and lepton transverse energy vectors respectively and $\slashed{\mathbf{E}}_T$ is the missing transverse energy vector.

The $W(\mu\nu)\gamma$ sample was selected from a trigger requiring an EM cluster with $E_T > 7$ GeV and a muon track with $p_T > 5$ GeV in the region $|\eta| < 1.7$. Kinematic selection was made requiring $p_T^\mu > 15$ GeV and $\not{E}_T > 15$ GeV. Muon tracks were required to satisfy standard muon identification requirements [12]. To reduce the background due to $Z(\mu\mu)\gamma$ production, events were rejected if they contained an extra muon track with $p_T > 8$ GeV.

The requirements on photons were common to both the electron and muon channel samples. We required $E_T^\gamma > 10$ GeV and the same geometrical and quality selection as for electrons, except that we required a tighter isolation, $I < 0.10$, and that there be no track matching the calorimeter cluster. The isolation parameter is defined as $I = [E(0.4) - EM(0.2)]/EM(0.2)$, where E(0.4) is total calorimeter energy inside a cone of radius $R$=0.4 and EM(0.2) is the electromagnetic energy in a cone of radius $R$=0.2. In addition, we required that the separation between the photon and lepton be $\Delta R_{l\gamma} > 0.7$. This requirement suppresses the radiative $W$ decay events as described above, and minimizes the probability for a photon cluster to merge with a nearby calorimeter cluster associated with an electron or a muon. The above selection criteria yielded 11 $W(e\nu)\gamma$ candidates and 12 $W(\mu\nu)\gamma$ candidates.

The backgrounds are from $W +$ jet(s) production, where a jet is misidentified as a photon; $Z(ll)\gamma$ production where one of the leptons from the $Z$ is undetected or mismeasured by the detector; and $W\gamma$ with $W \rightarrow \tau\nu$ followed by $\tau \rightarrow l\nu\bar{\nu}$. The first was determined from DØ data using the probability that a jet fakes a photon $P(j\rightarrow\gamma)$ which was estimated using multijet events (Fig. 6). The $Z\gamma$ background was estimated using the $Z\gamma$ event generator of Baur and Berger [10] and the background from $\tau$ decays was obtained using the ISAJET program [13]. The $W +$ jet(s) process was the dominant source of background in the electron channel while in the muon channel the $Z\gamma$ background was larger. This is due to the geometrical acceptance of the muon detector system and muon chamber inefficiencies.

|  | $W(e\nu)\gamma$ | $W(\mu\nu)\gamma$ |
|---|---|---|
| Selection Cuts: | | |
| $p_T^l$ | 25 | 15 |
| $\not{E}_T$ | 25 | 15 |
| $E_T^\gamma$ | 10 | 10 |
| Lepton $\eta$ | $|\eta|<1.1$ or $1.5<|\eta|<2.5$ | $|\eta|<1.7$ |
| Photon $\eta$ | $|\eta|<1.1$ or $1.5<|\eta|<2.5$ | |
| No. Candidates | 11 | 12 |
| Backgrounds: | | |
| $W + \text{jet(s)}$ | $1.7 \pm 0.9$ | $1.3 \pm 0.7$ |
| $Z\gamma$ | $0.11 \pm 0.02$ | $2.7 \pm 0.8$ |
| $W(\tau\nu)\gamma$ | $0.17 \pm 0.02$ | $0.4 \pm 0.1$ |
| Total | $2.0 \pm 0.9$ | $4.4 \pm 1.1$ |

Table 2. Summary of $W(e\nu)\gamma$ and $W(\mu\nu)\gamma$ selection, data and backgrounds.

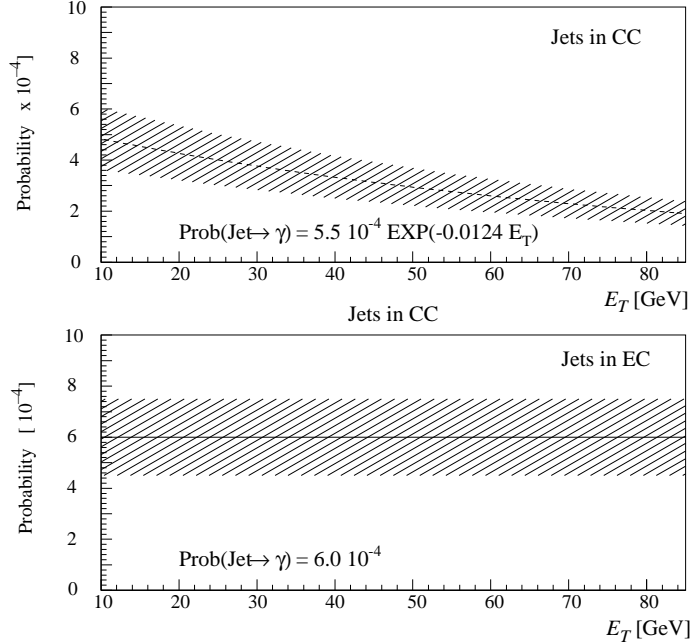

Fig. 6. Probability that a jet fakes a photon as a function of $E_T$ in the central calorimeter (top) and end calorimeters (bottom). The shadded band shows the estimated error.

The SM prediction for the number of events observed was obtained using the $W\gamma$ event generator of Baur and Zeppenfeld [14] together with a Monte Carlo simulation of the DØ detector. We calculated the cross section for $W\gamma$ production (with $E_T^\gamma > 10$ GeV and $\Delta R_{l\gamma} > 0.7$) to be $\sigma(p\bar{p} \to W\gamma) = 138^{+51}_{-38}\,(\text{stat}) \pm 21\,(\text{sys})$ pb compared with the SM prediction of $\sigma(p\bar{p} \to W\gamma) = 112 \pm 10$ pb. Therefore, the observed signal agrees with the SM prediction (within errors). Fig. 7 shows the kinematic distributions for the observed candidate events together with the SM signal prediction plus the sum of the estimated backgrounds. The shapes of these distributions show no deviations from the expectations. To set limits on the $WW\gamma$ vertex coupling parameters, we used a binned maximum likelihood fit to the photon $E_T$ distribution. Fig. 8 shows the 68% and 95% confidence level (CL) limits in the $\Delta\kappa^0$–$\lambda^0$ plane, where we have used a form factor scale of $\Lambda = 1.5$ TeV. Varying only one coupling at a time we find the following limits at the 95% CL:

$$-1.6 < \Delta\kappa^0 < 1.8 \quad \text{(for } \lambda^0 = 0\text{)}$$
$$-0.6 < \lambda^0 < 0.6 \quad \text{(for } \Delta\kappa^0 = 0\text{)}.$$

The possibility of a minimal U(1)-only coupling ($\kappa=\lambda=0$) indicated by the star in Fig. 8 is ruled out at the 88% CL.

The CDF $W\gamma$ analysis, described in [15], also sets limits on the $WW\gamma$ vertex couplings. These limits are shown in Fig. 9, together with the DØ limits described above.

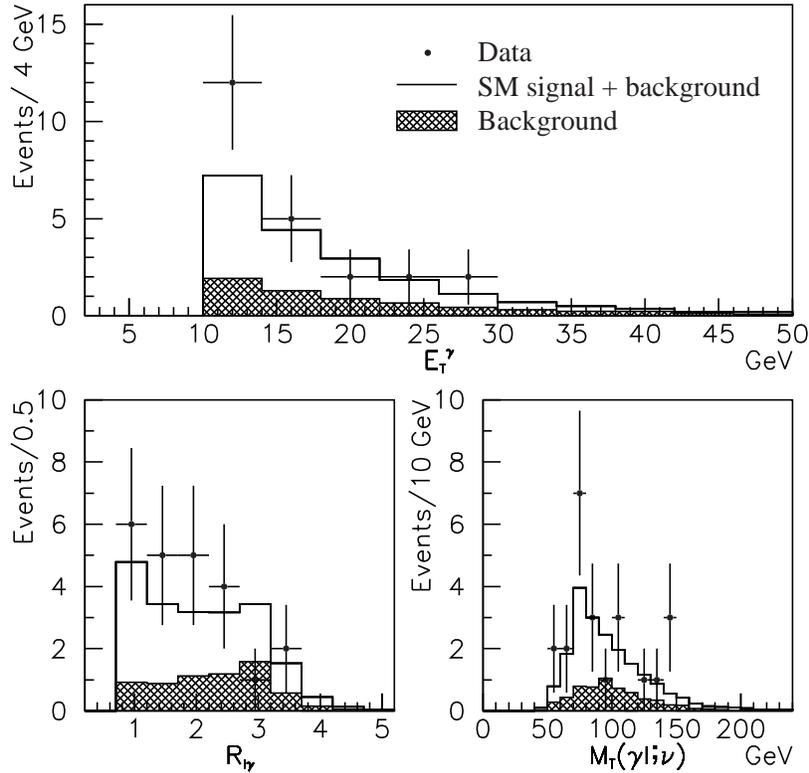

Fig. 7. Distribution of photon transverse energy $E_T^\gamma$ (top), $\Delta R_{l\gamma}$ (left) and transverse cluster mass $M_T(l\gamma;\nu)$ for the $W(e\nu)\gamma + W(\mu\nu)\gamma$ combined sample. The points are data. The shaded areas represent the estimated background and the solid histograms are the expected signal from the Standard Model plus the estimated background.

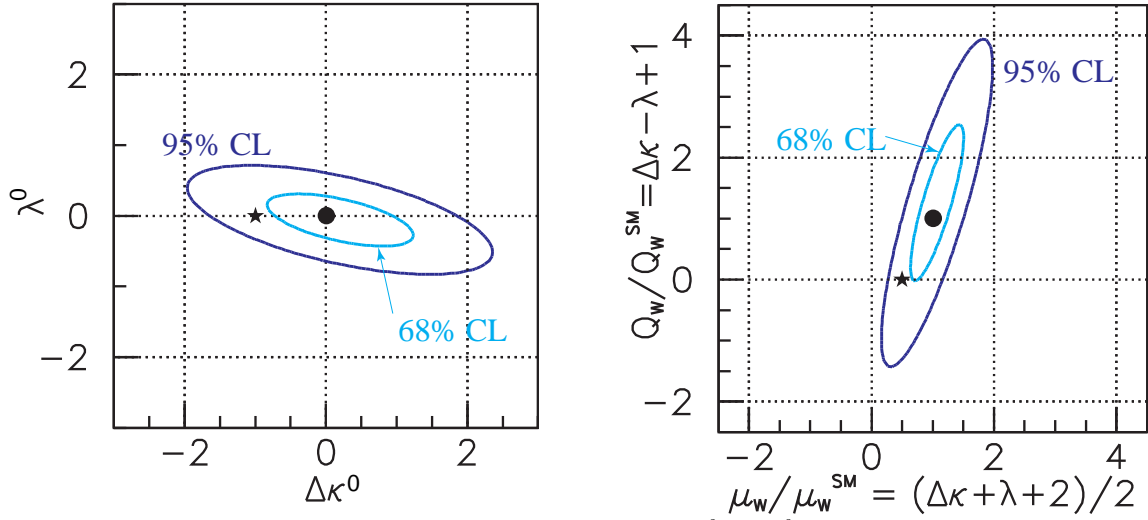

Fig. 8. Limits on CP conserving coupling parameters $\Delta\kappa^0$ and $\lambda^0$ (left) and on the magnetic dipole $\mu_W$ and electric quadrupole $Q_W$ moments. The ellipses represent the 68% and 95% CL exclusion contours. ● represents the Standard Model values, while ★ indicates the U(1)-only coupling of the W boson to the photon, $\Delta\kappa=-1$ and $\lambda=0$ ($\mu_W=e/2m_W$ and $Q_W=0$).

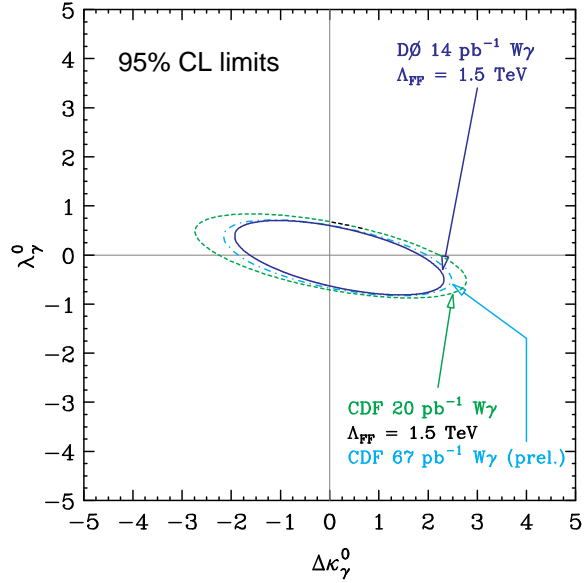

Fig. 9. Limits on the $WW\gamma$ anomalous coupling parameters from DØ and CDF.

## 5. $Z\gamma$ Production

At the Tevatron $Z\gamma$ production can be used to test for the possibility of non-zero $ZZ\gamma$ and $Z\gamma\gamma$ couplings. The SM $Z\gamma$ production processes are shown in Fig. 10 (a)-(c). These are analagous to the u- and t-channel and inner bremsstrahlung processes in $W\gamma$ prduction. For non-SM $ZZ\gamma$ or $Z\gamma\gamma$ couplings the diagram of Fig. 10 (d) results in an increased cross section and an enhacement of events with high-$E_T$ photons.

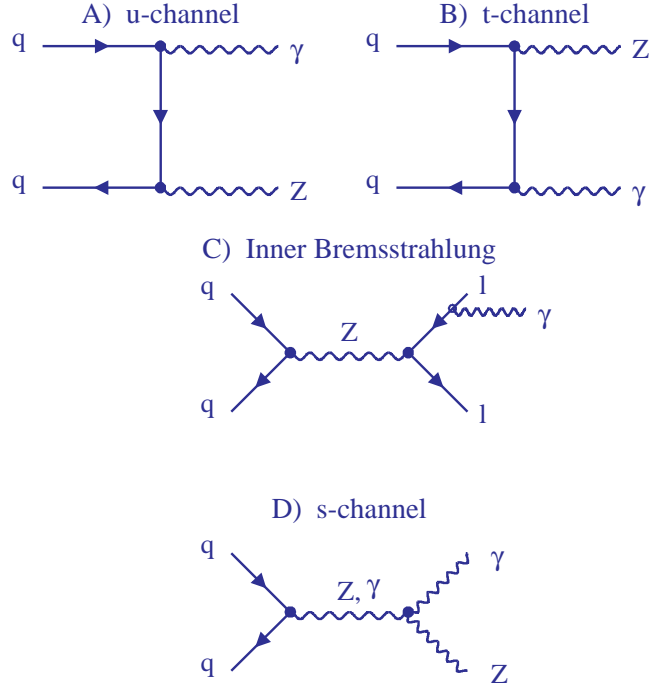

Fig. 10. Feynman diagrams contributing to the process $p\bar{p} \to l^+l^-\gamma$.

We searched for $Z\gamma$ events in which the Z decays into $e^+e^-$ or $\mu^+\mu^-$. The $ee\gamma$ sample was selected from a trigger requiring two isolated EM clusters, each with $E_T > 20$ GeV. One electron is required to pass the standard electron quality cuts and the second electron must pass all these cuts except the requirement of a matching track in the drift chambers. Both electrons were required to have $E_T > 25$ GeV. The photon must have $E_T > 10$ GeV and be separated from both electrons by $\Delta R_{l\gamma} > 0.7$. These selection criteria resulted in four $ee\gamma$ candidates and two $\mu\mu\gamma$ candidates.

The backgrounds were from (i) Z + jet(s) production where one of the jets fakes a photon or an electron (the latter case corresponds to the $ee\gamma$ signature if, additionally, one of the electrons from the $Z \to ee$ decay is not detected in a tracking chamber); (ii) QCD multijet production with jets being misidentified as electrons or photons; (iii) $\tau\tau\gamma$ production followed by the decay of each $\tau$ to $l\bar{\nu}_l\nu_\tau$. Backgrounds (i) and (ii) were calculated from data using the misidentification probabilities $P(\text{jet} \to e)$ and $P(\text{jet} \to \gamma)$ estimated from multijet events as described above. The $\tau\tau\gamma$ background was estimated using the ISAJET Monte Carlo event generator followed by a full GEANT simulation of the DØ detector.

After subtracting the backgrounds, the signal is $3.57^{+3.15}_{-1.91} \pm 0.06$ events for the $ee\gamma$ channel and $1.95^{+2.62}_{-1.29} \pm 0.01$ events for the $\mu\mu\gamma$ channel, where the first error is due to Poisson statistics and the second is due to the systematic uncertainty of the background estimate.

To set limits on anomalous $ZZ\gamma$ and $Z\gamma\gamma$ couplings we used a fit to the photon $E_T$ distribution, as described above for the $W\gamma$ analysis. Figure 11 shows the observed $E_T$ spectrum with the sum of the backgrounds plus SM signal. The latter was obtained using the Baur and Berger event generator [10] combined with a simulation of the DØ detector. The 95% CL limits on the CP-conserving anomalous coupling parameters $h^Z_{30}$ and $h^Z_{40}$, obtained by varying only one coupling at a time and using a form factor scale of $\Lambda = 500$ GeV are:

$ZZ\gamma$ coupling: $\quad -1.8 < h^Z_{30} < 1.8$
$\quad -0.5 < h^Z_{40} < 0.5;$

$Z\gamma\gamma$ coupling: $\quad -1.9 < h^\gamma_{30} < 1.9$
$\quad -0.5 < h^\gamma_{40} < 0.5.$

The 95% CL limit contour in the $h^Z_{30}$–$h^Z_{40}$ plane is shown in Fig. 12 together with the limits from CDF [16] and L3 [17]. The L3 measurement is based on an analysis of $Z \to \nu\bar{\nu}\gamma$. At the LEP energy the sensitivity to $h^\gamma_{40}$ is less than that to $h^Z_{30}$. At the Tevatron the anomalous couplings are probed at higher $\sqrt{\hat{s}}$, and because the helicity amplitude contributions grow like $(\sqrt{\hat{s}}/M_Z)^5$ for $h^\gamma_{40}$ and $(\sqrt{\hat{s}}/M_Z)^3$ for $h^Z_{30}$, the Tevatron measurements are more sensitive to $h^\gamma_{40}$.

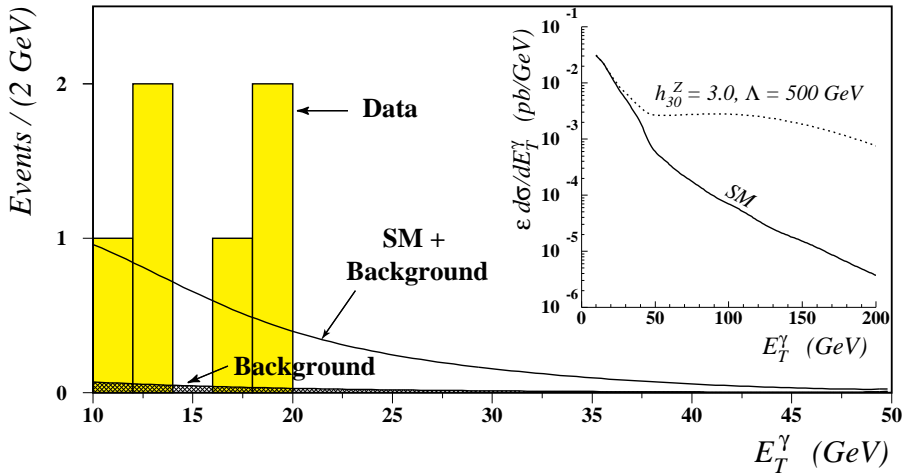

Fig. 11. Photon $E_T$ distribution for the $Z(ee)\gamma + Z(\mu\mu)\gamma$ candidate events. The sum of the backgrounds plus SM signal is also shown. The inset shows the enhancement of high $p_T$ photons for anomalous couplings.

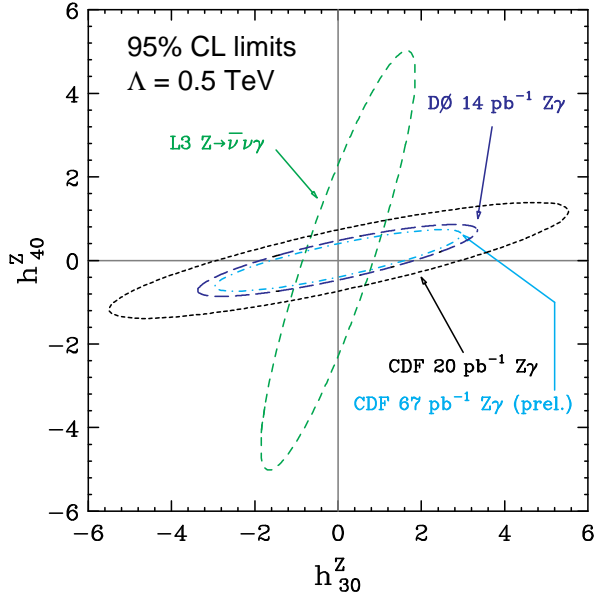

Fig. 12. Limits on the anomalous $ZZ\gamma$ coupling parameters from DØ, CDF and L3.

## 6. *WW* and *WZ* Production

We have searched for *WW* and *WZ* production using two decay modes: (a) $WW \to l\nu\, l\nu$, or the "dilepton" mode and (b) $WW, WZ \to l\nu\, q\bar{q}$, or the "lepton plus jets" mode. In the dilepton analysis [18] we search in the *ee*, $\mu\mu$ and $e\mu$ channels, requiring two high $p_T$ leptons and missing $E_T$ greater than 20 GeV. One event passes the selection criteria (an *ee* event). The SM prediction for the *WW* signal, based on a cross section of 9.5 pb [19] is 0.47 ± 0.07 events. We set an upper limit on the cross section for $p\bar{p} \to W^+W^- + X$ of 87 pb at the 95% CL, estimated based on one signal event including subtraction of the total estimated background of 0.56 ± 0.13 events.

The *W* pair production process is sensitive to the $WW\gamma$ and $WWZ$ couplings, since the s-channel propagator can be a $\gamma$ or *Z*. We set limits on the anomalous coupling parameters, assuming that $\Delta\kappa_Z = \Delta\kappa_\gamma$, $\lambda_Z = \lambda_\gamma$. The limits obtained from the obsereved cross section limit and the theoretical cross section as a function of $\Delta\kappa$ and $\lambda$ are shown in Fig. 14.

In the *preliminary* lepton plus jets analysis we search for $WW, WZ \to l\nu jj$ candidate events containing a $W \to e\nu$ decay ($E_T^e > 25$ GeV, $\not{E}_T > 25$ GeV forming a transverse mass > 40 GeV) and two jets consistent with $W \to jj$ or $Z \to jj$. The candidate events were required to contain at least two jets with $E_T^j > 25$ GeV and have a dijet mass of $50 < m_{jj} < 110$ GeV/c² (the largest mass was taken if the event had more than two jets). A cone algorithm with a radius $R = 0.3$ was used to reconstruct jets. This small cone size minimizes the probability for two jets from the *W* or *Z* boson to merge into one cluster in the calorimeter, especially for high $p_T$ *W* and *Z* bosons. Figure 13 shows the observed $p_T$ distribution of the $e\nu$ system. As shown, the data are dominated by background, mainly from (i) W+≥2jets events with $W \to e\nu$ and (ii) multijet production where one jet was misidentified as an electron and there was significant (mismeasured) missing $E_T$.

At large values of the $p_T(W \to e\nu)$ the backgrounds are relatively small and it is in this region where anomalous couplings would enhance the cross section, as shown in the lower plot of Fig. 13. Therefore we performed a binned likelihood fit to the $p_T^W$ spectrum above $p_T^W > 25$ GeV. The results, assuming the $WWZ$ and $WW\gamma$ coupling parameters are equal (see section 3), are shown in Fig. 14. At the 95% CL, for $\Lambda = 1.0$ TeV, we obtain

$$-0.9 < \Delta\kappa^0 < 1.1 \qquad \text{(for } \lambda^0 = 0\text{)}$$
$$-0.7 < \lambda^0 < 0.7 \qquad \text{(for } \Delta\kappa^0 = 0\text{)}.$$

Note that the limits on $\Delta\kappa$ from this channel are tighter than those from the $W\gamma$ channel, but note that we assumed $\Delta\kappa_Z = \Delta\kappa_\gamma$.

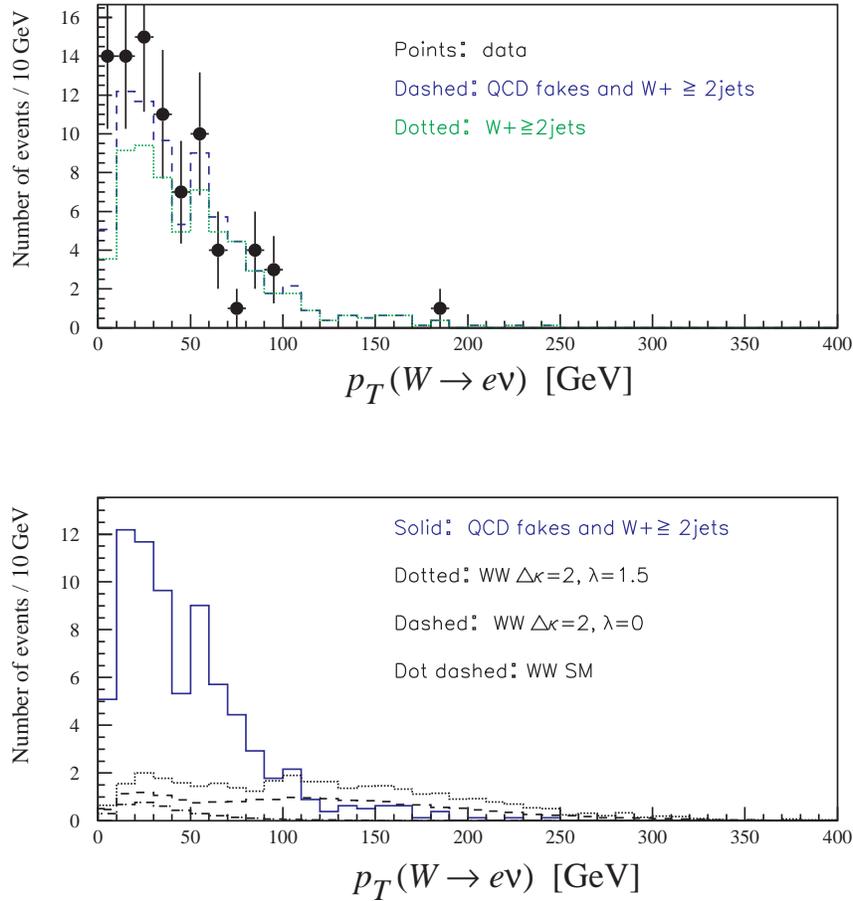

Fig. 13.  $p_T$ of the $e\nu$ system: data (points), $W + \geq 2$ jets background (dotted) and total background (dashed). In the second plot, the total background is compared with the predicted signal for anomalous couplings and shows the enhancement of the signal at high $p_T$.

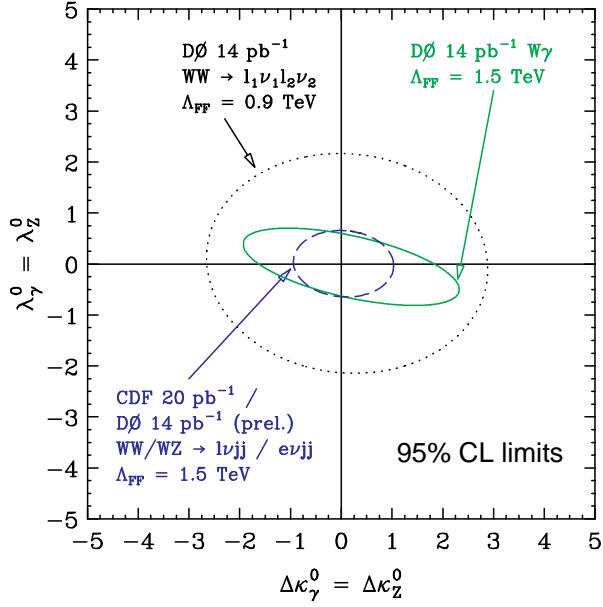

Fig. 14. Summary of limits on the anomalous $WWV$ coupling parameters $\Delta\kappa$ and $\lambda$ assuming $\Delta\kappa_Z = \Delta\kappa_\gamma$ and $\lambda_Z = \lambda_\gamma$.

## 7. Conclusions

Recent analyses of data based on the production of $W$ and $Z$ bosons using $\approx 12$ pb$^{-1}$ of data collected at the Fermilab Tevatron have resulted in considerable improvements in the measurement of the properties of the $W$ boson. As an example, the indirect measurement of the $W$ boson width is $\Gamma(W) = 2.062 \pm 0.059$ GeV (world average), corresponding to a precision of 3% on $\Gamma(W)$.

New studies of the physics of electroweak gauge boson pair production have also been undertaken. The production of $W\gamma$ and $Z\gamma$ events is in agreement with the predictions of the Standard Model, within the experimental sensitivity. These measurements allow limits to be set on anomalous electroweak gauge boson couplings. Searches for $WW$ and $WZ$ production are also used to set limits. The current 95% CL limits on the $WW\gamma$ coupling parameters from the $W\gamma$ channel are:

$$-0.9 < \Delta\kappa^0 < 1.1 \quad \text{(for } \lambda^0 = 0\text{)}$$
$$-0.7 < \lambda^0 < 0.7 \quad \text{(for } \Delta\kappa^0 = 0\text{)}.$$

and the limits on the $ZZ\gamma$ coupling parameters are ($\Lambda = 500$ GeV):

$$-1.8 < h^Z_{30} < 1.8 \quad \text{(for } h^Z_{40} = 0\text{)}$$
$$-0.5 < h^Z_{40} < 0.5 \quad \text{(for } h^Z_{30} = 0\text{)}.$$

It is important to note that the measurements described in this paper are based on Run 1a data only. Analyses of the data from the complete Run 1 ($\approx$100 pb$^{-1}$) will soon be completed and will lead to increases in the precision of these measurements. The upgraded DØ and CDF detectors will begin running at the Tevatron in 1999 collecting data sets of >2 fb$^{-1}$ in the first few years. As an example, the *W* width is expected to be measured with an error of $\pm$40 MeV from Run 1 and $\pm$20 MeV assuming an integrated luminosity of 10 fb$^{-1}$. With 10 fb$^{-1}$ of data the *WWV* vertex ($V=\gamma,Z$) coupling parameters can be tested at the level of $\approx$10%, and the *ZV$\gamma$* vertex couplings at the level of $10^{-2}$–$10^{-3}$. Complimentary tests of the anomalous couplings will also come from LEP II.

## Acknowledgments

I would like to thank my DØ colleagues, especially Liang-Ping Chen, Brajesh Choudhary, Marcel Demarteau, Steve Glenn, Mike Kelly, Greg Landsberg, Paul Quintas and Darien Wood. I also thank Ulrich Baur for helping me out on some of the theoretical issues.